\documentclass[prd]{revtex4}%
\usepackage{caption}
\usepackage{amsmath}
\usepackage{graphicx}
\usepackage{epsfig}
\usepackage{dcolumn}
\usepackage{bm}
\usepackage{slashed}
\usepackage{amsfonts}
\usepackage{amssymb}
\usepackage[english]{babel}
\usepackage{graphicx}
\usepackage{wrapfig}%
\providecommand{\U}[1]{\protect\rule{.1in}{.1in}}
\begin{document}
\title{Real scalar field stars of the EKG equations including matter }
\author{A. Cabo Montes de Oca and D. Suarez Fontanella }
\affiliation{\textit{Theoretical Physics Department, Instituto de Cibern\'{e}tica,
Matem\'{a}tica y F\'{\i}sica, Calle E, No. 309, Vedado, La Habana, Cuba. }}

\begin{abstract}
\noindent Static (not stationary) solutions of the Einstein-Klein-Gordon (EKG)
equations including matter are obtained for real scalar fields. The scalar
field interaction with matter is considered. The introduced coupling allows
the existence of static solutions in contraposition with the case of the
simpler EKG equations for real scalar fields and gravity. Surprisingly, when
the considered matter is a photon-like gas, it turns out that the
gravitational field intensity at large radial distances becomes nearly a
constant, exerting an approximately fixed force to small bodies at any
distance. The effect is clearly related with the massless character of the
photon-like field. It is also argued that the gravitational field can generate
a bounding attraction, that could avoid the unlimited increase in mass with
the radius of the obtained here solution. This phenomenon, if verified, may
furnish a possible mechanism for explaining how the increasing gravitational
potential associated to dark matter, finally decays at large distances from
the galaxies. A method for evaluating these photon bounding
effects is just formulated in order to be further investigated.

\end{abstract}
\maketitle

\section{Introduction}

The search for boson and fermion-boson stars is a subject of current interest
in modern Astrophysics \cite{jetzer,liddle1,liddle2,liddle3,urena}. The interest in
the theme is intensified the circumstances related with the search for purely
boson stars. As it is known, the Einstein-Klein-Gordon (EKG) equations have no
static solutions for real scalar fields \cite{jetzer}. Also, although the
complex EKG equations show centrally symmetric solutions, they need to be time
dependent in a stationary form (harmonic). In the classic references on
fermion-boson stars \cite{liddle1,liddle2,liddle3,urena} it was investigated the
existence and stability of such systems. The discussion in these works defined
conditions for the existence of such objects for a general case, but in which
still the interactions between the scalar fields and matter where not considered.

In the present work, we relax the assumption of the lack of interaction
between the scalar field and matter for the considered systems. This is done
in order to inspect the possibility for the appearance of solutions in which
the scalar field becomes static, that is, time independent when real scalar
field are considered. In order to simplify the discussion, we write the EKG
equations by considering matter described by simple constituent relations
including the photon gas one. The interaction between the scalar field and the
gas is introduced by assuming that the scalar field source is proportional to
the matter energy density.

Firstly, it is considered a matter being close to the limit of pressureless
gas. The initial conditions for the energy density at the coordinate origin,
for a centrally symmetric solution was then fixed to a given value. Further,
the equations were solved by specifying a tentative scalar field value at the
origin. This first step solutions was then inspected for the behavior of the
scalar field at large radial distance. This behavior emerged in only two
types: a) One in which the scalar field tends to be singular an positive at
some radial distance, and b) Another, in which the field tended to be also
singular but negative at some radius value. Then, we noted that, if the field
tends to be positive (negative) valued, the reduction (increasing) of its
initial value at origin, reduced the tendency to positive (negative) values,
by at the same time always augmenting the radius value at which the field
become singular. Therefore, after properly selecting the initial values by
iterating the above described process, a solution in which the scalar field
attains a Yukawa like behavior at infinity can be approached. The resulting
energy and pressure of the matter became centered in a bounded spherical
vicinity of the origin. As for the temporal $v$ and the radial $u$ diagonal
components of the metric and its inverse, respectively, they both approach
constants at large distances, thus reproducing the Minkowski space-time. At
short distances the temporal metric defines a gravitational potential which
bounds the matter to the origin, at which its minimum sits. These results
indicate the existence of static solutions of the EKG equations in presence of
matter, when it interacts with the scalar field.  The stability of the solutions is preliminarily 
 investigated. For this purpose, a usual stability  criterion for simpler stars constituted by fluids 
  is verified to be valid \cite{teukolsky}.

In second place, we also examined the solution associated to a matter related
to a photon-like gas. The procedure for obtaining the solution was identical.
As before, after fixing the matter density and the scalar field at the origin,
the before described process to define a decreasing scalar field at large
distances became closely similar. However, a different outcome arose in
connection with the temporal component of the metric at large distances. It,
surprisingly got a nearly linear radial dependence at large distance, in place
of the expected constant behavior associated to the Minkowski space-time. This
potential behavior corresponds to a approximately constant gravitational force
over any small body at large distances. Such a behavior should be associated
to the special massless character of the photon-like field, which might not
allow the gravitational forces to fully confine the massless particles as they
bound massive ones.

The above results strongly motivates to consider a photon-like field to
describe the velocity $vs$ radius curves of the galaxies, being associated to
dark-matter. In addition, although it not corresponds to the main current
point of view that real photons can act as dark-matter (\cite{peebles}%
,\cite{rees}), we also examine this question, seeking for unnoticed possibilities.

In a preliminary study of these issues, we considered the determination of the
parameters of the considered star, in order to reproduce the experimentally
observed velocity $vs$ radius curves. For this purpose the mass of the scalar
field was fixed to a value defining a galaxy of a typical size. The rotation
curve following was of the type C of the three A,B and C kinds in which the
galaxies rotation curves are classified \cite{classif}. That is, the velocity
curve after rapidly growing for small radial distances, continue to grow, but
with a smaller slope.  Then, the value of the photon like particles energy
density at the radial distances near the outside of the galaxy was evaluated
and compared with the CMB energy density. The obtained result for the energy
density at such points resulted enormously higher than the CMB energy density.
Therefore, this first checking of the possibility that dark-matter were
constituted by real photons and not dark-matter analogous massless particles
gave a negative result.

However, some causes for the discrepancy of the result for the energy density
can be thought. One them, is that EKG equations solved here consider the
photons as obeying a free dispersion curve, which leads to the employed
equation of state: $\epsilon=3 \, p$. In this sense, it can be noted that the
photons under the gravitational attraction, can be imagined to acquire some of
the properties of confined photons in a resonant cavity. By example a discrete
spectrum, that could have similar effects as a mass for photons. In order to
discuss this possibility a simple model is presented in which a Newtonian
gravitational potential is able to create mass like terms in the Maxwell
equations. They, explicitly makes the photon wave to decrease in the direction
in which the potential increases.  In a coming work, we expect to consider a
classical statistical model for photons in each space-time point, but which
statistics will be controlled by the local values of the metric. This
consideration, we expect that can introduce an attracting action of the
gravitational field upon the photon density. This effect could bound them, by
consequently stopping the growing of the potential and leading to an
asymptotic Minkowski space. Upon this, it could be the case that the energy
density of the modified galaxy solution can perhaps tend to the CMB density
far form the galaxy \cite{massive}.

We also examined the bounds posed on the real photons acting as dark-matter,
by the fact that the photon matter should not be observable. In this sense,
the Tolman theorem expressing the temperature a black-body radiation subject
to gravitational field becomes helpful. Let us assume as above, that photons
are able to be trapped in galaxies by the mentioned self-consistent effects.
Then, the natural value of the radiation temperature in external zones to the
galaxies is the Cosmic-Microwave-Background one $T_{cmb}$. Therefore, from
Tolman theorem it follows that no matter the high temperatures the photon
radiation can attain at some interior point of the structure, the radiation
coming to the Earth from that point, will always have nearly 3 Kelvin degrees
of temperature. Thus, it could not be easily observed. This is a conclusion
not ruling out the possibility for the real photons playing the role of
dark-matter. A deeper investigation should however be pursued in order to
check, if there is still some way for the CMB could be in thermal equilibrium
with an internal to the  galaxy photon gas defining the halo. It should be
recognized that the enormous difference between the energy densities  allowing
the photons to reproduce the rotation curves, and the CMB energy density
suggests that the answer should be negative.  However, the mechanism discussed
here has not clear restrictions to work if a photon like but really dark
matter is constituting the dark matter.

In section 2, the Lagrangian of the system is presented and the equations for
the time independent spherically symmetric solutions written. Section 3,
exposes the derivation of the static solution for nearly pressureless matter.
Next, section 4, is devoted to describe the second solution associated to the
photon-like matter. Section 5, considers the preliminary inspection of the
possibility that the second solution might describe the velocities $vs$ radius
curves in galaxies. The conclusions resume the content of the work and
describe possible extensions.

\section{The field equations}

Let us consider the metric defined by the following squared interval and
coordinates%
\begin{align}
ds^{2}  &  =\mathit{v}(\rho){dx^{o}}^{2}-u(\rho)^{-1}d\rho^{2}-\rho
^{2}(sin^{2}\theta\text{ }d\varphi^{2}+d\theta^{2}),\\
x^{0}  &  =c\text{ }t\text{, \ \ \ }x^{1}=\rho,\\
x^{2}  &  =\varphi,\text{ \ }x^{3}\equiv\theta,
\end{align}
where the CGS unit system is employed. Therefore, the Einstein tensor
$G_{\mu\nu}$ components in terms of functions $u,v$ and the radial variable
$\rho$ are evaluated in the form
\begin{align}
{G_{0}^{0}}  &  ={\frac{u^{\prime}}{\rho}}-{\frac{1-u}{\rho^{2}}},\\
G_{1}^{1}  &  ={\frac{u}{v}}{\frac{v^{\prime}}{\rho}}-{\frac{{1-u}}{\rho^{2}}%
},\\
{G_{2}^{2}}={G_{3}^{3}}  &  ={\frac{u}{2v}}{v^{\prime\prime}}+{\frac
{uv^{\prime}}{4v}(\frac{u^{\prime}}{u}-\frac{v^{\prime}}{v})}\nonumber\\
&  +{\frac{u}{2\rho}(\frac{u^{\prime}}{u}+\frac{v^{\prime}}{v})}.
\end{align}

The physical system interacting with gravity that will be will considered is
composed of a scalar field and gas of matter. The scalar field will be assumed
to also linearly interact with an external source associated to it. The action
of the field will have the form%
\begin{equation}
{S}_{mat-\phi}=\int L\sqrt{-g} \,\, d^{4}x,
\end{equation}
with a Lagrangian density given by
\begin{equation}
{L}={\frac{1}{2}}(g^{\alpha\beta}{\Phi}_{,\alpha}{\Phi}_{,\beta}+m^{2}{\Phi
}^{2}+2\text{ }J(\rho)\text{ }\Phi), \label{denslag}%
\end{equation}
in which the $A_{,\alpha}$ mean the derivative of $A$ over the variable
$\alpha$.

This Lagrangian determines an energy momentum of the form
\begin{equation}
(T_{mat-\phi})_{\mu}^{\nu}=-\frac{\delta_{\mu}^{\nu}}{2}(g^{\alpha\beta}{\Phi
}_{,\alpha}{\Phi}_{,\beta}+m^{2}{\Phi}^{2}+2\text{ }J(\rho)\text{ }\Phi),
\end{equation}
which afterwards can be added to the energy momentum tensor of the matter
(\cite{Weinberg}):
\begin{equation}
(T_{e,P})_{\mu}^{\nu}=P\,\delta_{\mu}^{\nu}+u^{\nu}u_{\mu}(P+e),
\end{equation}
to write the total energy momentum tensor as%
\begin{align}
T_{\mu}^{\nu}  &  =-\frac{\delta_{\mu}^{\nu}}{2}(g^{\alpha\beta}{\Phi
}_{,\alpha}{\Phi}_{,\beta}+m^{2}{\Phi}^{2}++2\text{ }J(\rho)\text{ }%
\Phi)\nonumber\\
&  \text{ \ \ \ }+g^{\alpha\nu}{\Phi}_{,\alpha}{\Phi}_{,\mu}+P\text{ }%
\delta_{\mu}^{\nu}+u^{\nu}u_{\mu}(P+e).
\end{align}

Since static field configurations are being searched, the four-velocity
reduces to the form in the local rest system%
\begin{equation}
u^{\mu}=(1,0,0,0).
\end{equation}

After these definitions, the Einstein equations can be written as follows
\begin{equation}
G_{\mu}^{\nu}=\kappa\,\,\hspace{0.1mm}T_{\mu}^{\nu}, \label{Einstein}%
\end{equation}
in which both of the tensors appearing are diagonal and the gravitational
constant has the value%
\begin{equation}
\kappa=8\pi\text{ }l_{P}^{2},
\end{equation}
in terms of the Planck length $l_{P}=1.61\times10^{-33}$ cm. Their explicit
forms are
\begin{align}
{\frac{u^{\prime}}{\rho}}-{\frac{1-u}{\rho^{2}}}  &  =-\kappa\text{ }[\frac
{1}{2}(u\Phi_{,\rho}^{2}+m^{2}\Phi^{2}+2\text{ }j\text{ }\Phi)+e],\\
{\frac{u}{v}}{\frac{v^{\prime}}{\rho}}-{\frac{{1-u}}{\rho^{2}}}  &
=\kappa\text{ }[\frac{1}{2}(u\Phi_{,\rho}^{2}-m^{2}\Phi^{2}-2\text{ }j\text{
}\Phi)+P\,],\\
{\frac{\rho^{2}u}{2}}{v^{\prime\prime}}+\frac{\rho^{2}}{4}u\text{ }v^{\prime
}({\frac{{u}^{\prime}}{u}-}\frac{{v}^{\prime}}{v})+\frac{\rho}{2}u\text{
}v^{\prime}  &  =\kappa\text{ }[\frac{1}{2}(u\Phi^{\prime2}+m^{2}\Phi
^{2}+2\text{ }j\text{ }\Phi)+P\,],\\
{\frac{\rho^{2}u}{2}}{v^{\prime\prime}}+\frac{\rho^{2}}{4}u\text{ }v^{\prime
}({\frac{{u}^{\prime}}{u}-}\frac{{v}^{\prime}}{v})+\frac{\rho}{2}u\text{
}v^{\prime}  &  =\kappa\text{ }[\frac{1}{2}(u\Phi^{\prime2}+m^{2}\Phi
^{2}+2\text{ }j\text{ }\Phi)+P\,].
\end{align}

These are four equations, the last two of which are identical. Thus, there are
three independent Einstein equations in the problem. But, the third and the
equivalent fourth ones can be substituted by a simpler relation. It comes from
the Bianchi identities (\cite{Weinberg}):
\begin{equation}
G_{\mu\text{ };\text{ }\nu}^{\nu}=0, \label{Bianchi}%
\end{equation}
where the semicolon indicates the covariant derivative of the tensor
$G_{\mu\text{ }}^{\nu}.$ After assuming the satisfaction of the Einstein
equations (\ref{Einstein}) the $G_{\mu}^{\nu}$ tensor in (\ref{Bianchi}) can
be substituted by the energy momentum tensor $T_{\mu}^{\nu}$ leading to the
relation
\begin{equation}
-\Phi\text{ }J^{\prime}+P^{\prime}+\frac{v^{\prime}}{2v}(P+e)=0.
\end{equation}
This is a dynamic equations for the energy, the pressure and the scalar field,
substituting the two equivalent Einstein equations being associated to the
both angular directions.

The last of the equations of movement for the system is the Klein-Gordon one
for the scalar field. It can be obtained by imposing the vanishing of the
functional derivative of the action $S_{mat-\phi}$ with respect to the scalar
field
\begin{align}
\frac{\delta S_{mat-\phi}}{\delta\Phi(x)}  &  \equiv\frac{\partial}{\partial
x^{\mu}}\frac{\partial L}{\partial\Phi_{,\mu}}-\frac{\partial L}{\partial\Phi
}\nonumber\\
&  \equiv\frac{1}{\sqrt{-g}}\frac{\partial}{\partial x^{\mu}}(\sqrt{-g}%
g^{\mu\nu}\Phi_{,\nu})-m^{2}\Phi-J\nonumber\\
&  =0,
\end{align}
a relation that after employing the temporal and radial Einstein equations in
\ (\ref{Einstein}) can be rewritten in the form
\begin{align}
J(\rho)+m^{2}\Phi(\rho)-u(\rho)\text{ }\Phi^{\prime\prime}(\rho)  &
=\Phi^{\prime}(\rho)(\frac{u(\rho)+1}{\rho}-\rho\text{ }\kappa\text{ }%
(\frac{m^{2}\Phi(\rho)^{2}}{2}+\nonumber\label{escalar}\\
&  J(\rho)\text{ }\Phi(\rho)+\frac{e(\rho)-P(\rho)}{2}))
\end{align}

Therefore, the three relevant for the problem EKG equations are resumed as
\begin{align}
{\frac{u^{\prime}(\rho)}{\rho}}-{\frac{1-u(\rho)}{\rho^{2}}}  &  =-\text{
}\kappa\text{ }[\frac{1}{2}(u(\rho)\Phi^{\prime2}(\rho)+m^{2}\Phi(\rho
)^{2}+2\text{ }J(\rho)\Phi(\rho))+e(\rho)],\\
{\frac{u(\rho)}{v(\rho)}}{\frac{v^{\prime}(\rho)}{\rho}}-{\frac{{1-u}(\rho
)}{\rho^{2}}}  &  =\kappa\text{ }[\frac{1}{2}(u(\rho)\Phi^{\prime2}%
(\rho)-m^{2}\Phi(\rho)^{2}-2\text{ }J(\rho)\text{ }\Phi(\rho))+p(\rho)\,],\\
J(\rho)+m^{2}\Phi(\rho)-u(\rho)\text{ }\Phi^{\prime\prime(\rho)}  &
=\Phi^{\prime}(\rho)(\frac{u(\rho)+1}{\rho}-\rho\text{ }\kappa\text{ }%
(\frac{m^{2}\Phi(\rho)^{2}}{2}+J\text{ }(\rho)\Phi(\rho)+\frac{e(\rho
)-p(\rho)}{2})).
\end{align}

In order to work with dimensionless forms of the equations, let us define the
new radial variable, scalar field and parameters as follows
\begin{align}
r  &  =m\rho,\\
\phi(r)  &  =\sqrt{8\pi}l_{p}\Phi(\rho),\\
j(r)  &  =\frac{\sqrt{8\pi}l_{p}}{m^{2}}J(\rho),\\
\text{\ \ }\epsilon(r)  &  \equiv\frac{8\pi l_{p}^{2}}{m^{2}}e(\rho),\text{
\ \ \ }\\
p(r)  &  =\frac{8\pi l_{p}^{2}}{m^{2}}P(\rho).
\end{align}
It should noted that the new variable $r$ has dimension of $gr \times cm$ .

Therefore, the to be worked EKG equations in the new coordinates will be
\begin{align}
{\frac{u^{\prime}(r)}{r}}-{\frac{1-u(r)}{r^{2}}}  &  =-\frac{1}{2}%
(u(r)\phi^{\prime}{(r)}^{2}+\phi(r)^{2}+2j(r)\phi(r))-\epsilon
(r),\label{eecuaadim1}\\
\frac{u(r)}{v(r)}\frac{v^{\prime}(r)}{r}-{\frac{1-u(r)}{r^{2}}}  &  =-\frac
{1}{2}(-u(r){\phi}^{\prime}{(r)}^{2}+\phi(r)^{2}+2j(r)\phi
(r))+p(r),\label{eecuaadim2}\\
(\epsilon(r)+p(r))\frac{v^{\prime}(r)}{2v(r)}-\phi(r)\text{ }j^{\prime}(r)  &
=0,\label{eecuaadim3}\\
j(r)+{\phi}(r)-u(r)\text{ }{\phi}^{\prime\prime}(r)  &  ={\phi}^{\prime
}(r){\Large (}\frac{u(r)+1}{r}-r\text{ }(\frac{{\phi}(r)^{2}}{2}+j(r){\phi
}(r)+\frac{\epsilon(r)-p(r)}{2}){\Large )} . \label{eecuaadim4}%
\end{align}

Note that in order to simplify the notation, the same letter $u$ and $v$ had
been used to indicates the metric components in the new variables. That is, we
will write $u(r)=u(\rho)$ and $v(r)=v(\rho)$ in spite of the fact that
functional forms of the two quantities can not be equal. This should not
create confusion.

\subsection{ The constitutive relations and the matter-field interaction}

Let us now define the functional form of the constitutive relation for matter
as expressed in the new coordinate system
\begin{equation}
\epsilon(r)=n\text{ }p(r),
\end{equation}
where the parameter $n$ defines the ratio between the energy density and the
pressure. This simple form allows to consider the cases of nearly pressureless
dust for $n$ large and the case of a photon gas for $n=3.$

As mentioned before, the discussion will consider the interaction between
matter and the scalar field. It will be taken into account by assuming the
source of the scalar field $j(r)$ becomes proportional to the matter energy
$\epsilon(r):$%
\begin{equation}
j(r)=g\text{ }\epsilon(r).
\end{equation}
It should be stressed that we have considered the matter in a simpler approach
with respect to the more detailed one employed in references
\cite{liddle1,liddle2}. This was done in order to concentrate the discussion
in the relevant issue addressed in this paper: the role of the inclusion of
matter scalar field interaction.

\section{The EKG including matter static scalar field  solution}

Once the physical system had been defined, let us present in this section a
particular kind of solutions of the EKG equations for a real scalar field
interacting with matter. The objective of the presentation is to show the
existence of static configurations of a centrally symmetric star formed by
scalar field and matter contents. The found solution tends to reproduce the
Schwarzschild space-time with a Yukawa like scalar field decaying at large
distances \cite{jetzer}.

We start to search for it by fixing initial conditions for the radial evolution
defined at very small radial distance $\delta=10^{-6}$
\begin{align}
u(\delta) &  =1,\\
v(\delta) &  =1,\\
\phi(\delta) &  =\phi_{0}=0.65,\\
p(\delta) &  =p_{0}=0.0595725.
\end{align}
These initial conditions were fixed not exactly at the coordinate axis, but at
a very small non vanishing radial distance $\delta$, because the equations for
$u(r)$ and $v(r)$ are singular at $r=0.$ This singularity also enforces the
value of $u$ to tend to one in the limit $r->0$ assumed that the solution for
this quantity is regular. To solve the equations we programmed the
differential equations by using the software Mathematica. The proportionally
constant between the scalar field source and the energy density of the matter
was chosen as%
\[
g=0.9.
\]
\begin{figure}[h]
%[h]
\par
\begin{center}
\hspace*{-0.5cm}\includegraphics[width=70mm]{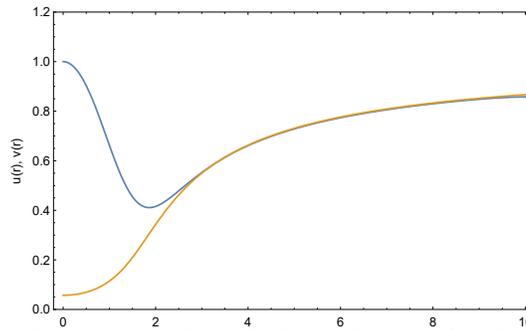}
\vspace{-1cm}
%\end{center}
\end{center}
\caption{ The figure shows the evolution with the radial coordinate $r$ for
the two fields $u(r)$ and $v(r)$. Note that at large radial distances the
metric components coincide, which indicates that the metric tensor tends to be
the Minkowski one, faraway from the symmetry center. At small distances at
which the matter and field energy densities start to grow, $u$ and $v$
deviates one from another: $u$ tends to the unit at the origin and $v$ reduces
its value to a minimum at this point. Note that the trapping of the matter
near the origin is compatible with the interpretation of $v$ as a
gravitational potential, which attracts matter to the region in which its
minimum appears. }%
\label{uv}%
\end{figure}

The procedure for obtaining the solutions was as follows. Firstly, it was
fixed the specified value for the pressure at the point $r=\delta.$ The
required value of $u(\delta)=1$ was imposed. Arbitrarily we fixed
$v(\delta)=1.$ This arbitrary character of the initial condition for
$v(\delta)$ is associated to the fact that for any solution of the general
equations, the multiplication of the function $v(r)$ by an arbitrary constant
is also a solution. Thus $v(\delta)=1$ can be always chosen. The employed
property directly follows from the fact that the EKG equations only depend on
$v(r)$ through the ratio $\frac{v^{\prime}(r)}{v(r)}$. \begin{figure}[h]
\begin{center}
\hspace*{-0.7cm}\includegraphics[width=70mm]{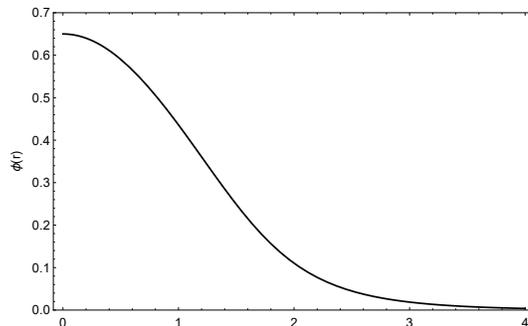} \vspace{-1cm}
\end{center}
\caption{ The plot show the radial behavior of the scalar field. In the large
radial distance region, in which the equations become linear ones, the scalar
field gets a decaying value corresponding with the Yukawa solution. }%
\label{phi}%
\end{figure}

Afterwards, in a first step, the equations were solved by fixing an arbitrary
value of the scalar field near the origin $\phi(\delta)=\phi_{0}$. The
possible results in this first stage were two fold: a first one in which the
scalar field grew to positive singular value at a given radial distance; a
second one in which the singular values resulted to be negative also at a
specific radial distance. Then, it was possible to note that decreasing
(increasing) the value of $\phi_{0},$ reduced the positive (negative) singular
values, while at the same time, in both cases the radial distance at which the
singularity appears increased. Repeating iteratively this process of
adjustment of the values of $\phi_{0},$ the singularity position was fixed
each time at larger distances. After this, it became clear that the scalar
field solution tended to reproduce the small field Yukawa like solution of the
Klein-Gordon equation. Upon this, the matter and scalar field energy densities
and pressures became concentrated in vicinity of the origin.

The figure \ref{uv} shows the values of the metric components. As it can be
noticed, at large distances from the symmetry center, the metric tends to be
the flat space Minkowski one. It should be here remarked that the solution for
$v (r)$ was multiplied by a constant in order to enforce the equality between
$u$ and $v$ at large distance, which reproduces the Minkowski space in the
faraway region. At small distances, the value of $u$ starts deviating from the
value of $v$. This occurs in the region in which the matter and scalar field
densities are mainly concentrated. While $u$ tends to the unit in approaching
the symmetry point, the temporal component of the metric tends to a minimum
value. This should be the case, if this metric component takes the role of
gravitational potential attracting the matter to the zone in which its minimum
value appear.

The resulting solution for the scalar field is shown in figure \ref{phi}. Far
form the center, the behavior is exponential as it should be, because in
Minkowski space the only real decaying radially symmetric solution of the
Klein-Gordon (KG) equation is the Yukawa potential one.

\begin{figure}[h]
\begin{center}
\includegraphics[width=60mm]{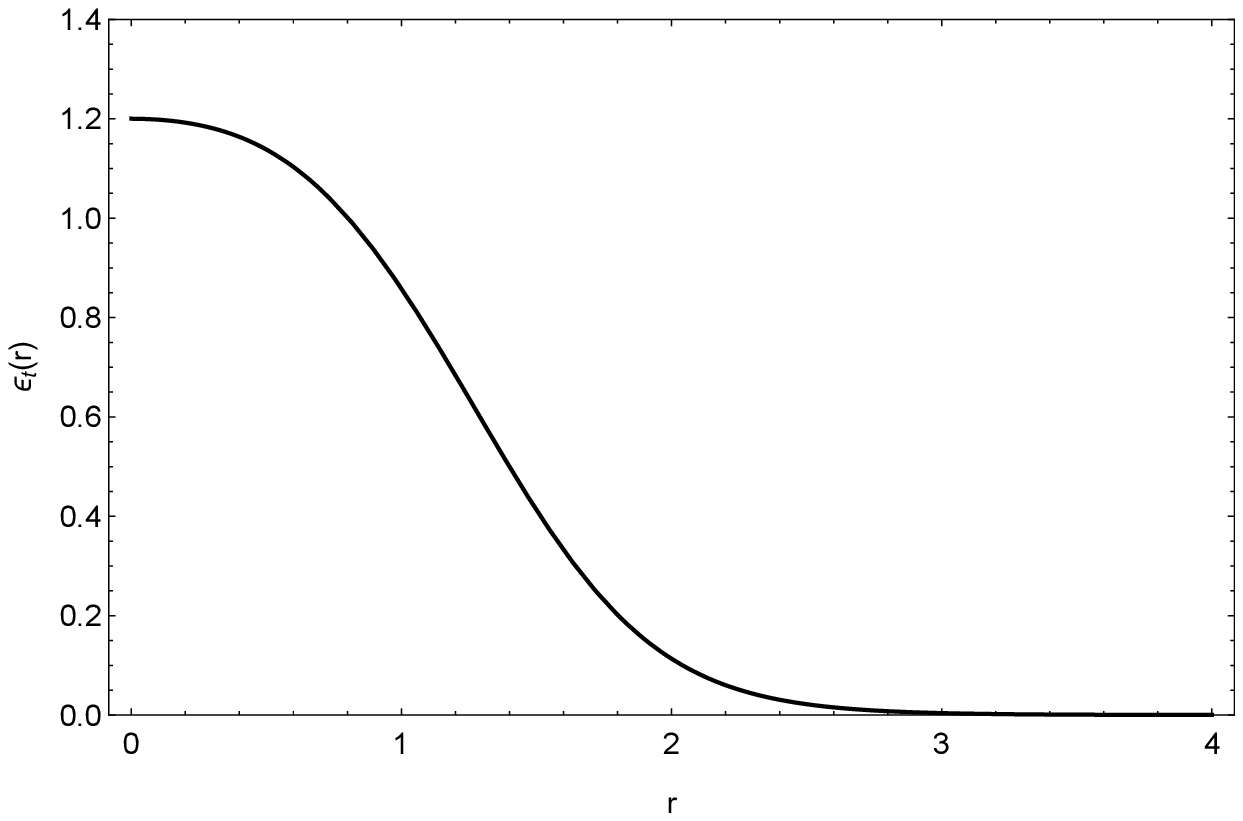} \hspace{1cm}%
\includegraphics[width=60mm]{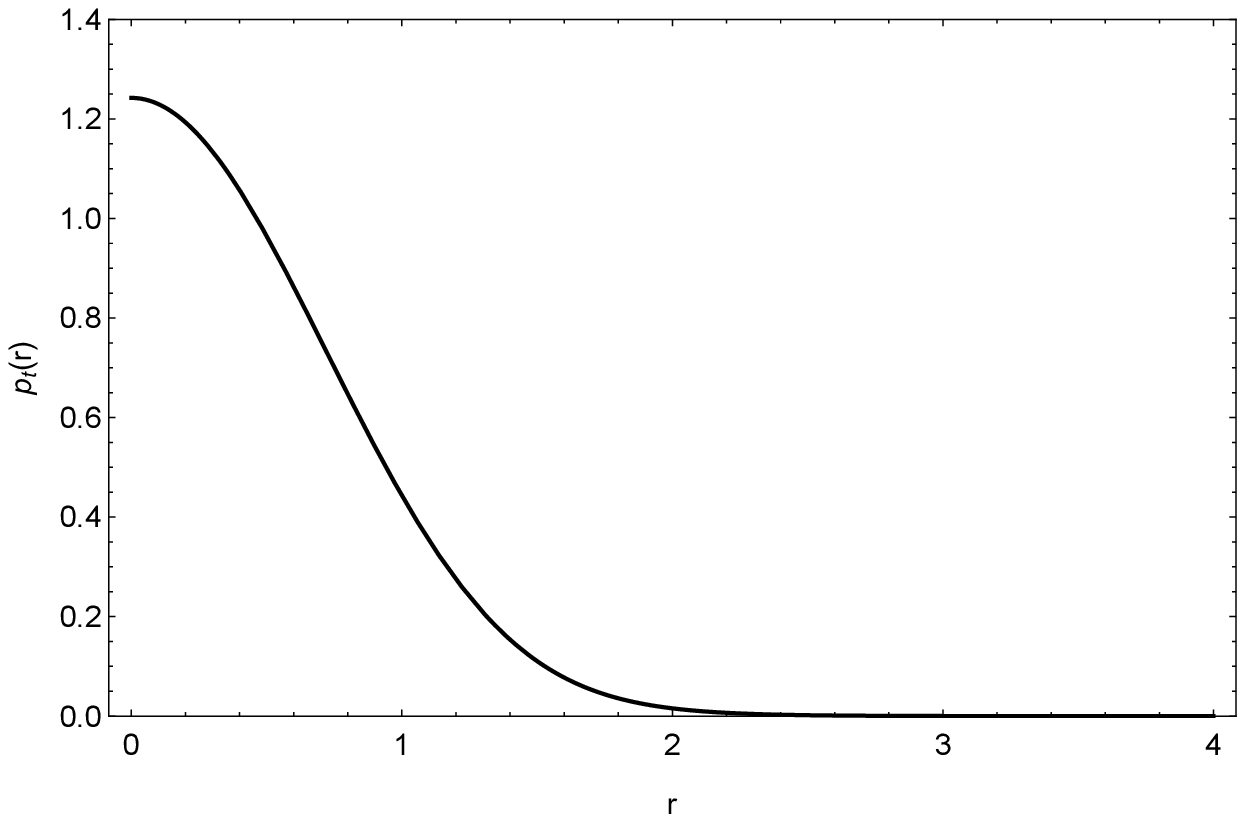}
\end{center}
\caption{ The two plots show the radial behavior of the total energy density
and the total pressure of the solution. The central role of the scalar
field-matter interaction in allowing the solution to exist, is compatible with
the fact that the matter energy density closely overlaps with the scalar field
one. }%
\label{etpt}%
\end{figure}

As for the total energy and pressure, defined as%
\begin{align}
\epsilon_{t}(r) &  =\frac{1}{2}(u(r)\phi^{\prime}{(r)}^{2}+\phi(r)^{2}%
+2j(r)\phi(r))+\epsilon(r),\\
p_{t}(r) &  =\frac{1}{2}(u(r)\phi^{\prime}{(r)}^{2}-\phi(r)^{2}-2j(r)\phi
(r))+p(r),
\end{align}
they show to be concentrated in spatial regions being similar in size to the
ones in which the scalar field energy is localized. Their radial dependence
are shown in figure \ref{etpt}. This property is compatible with the
determining role of the matter-field interaction in allowing the existence of
the solution.

\subsection{Stability analysis }

The full stability analysis of the considered physical system including a scalar field, should follow from
the general linearized equations of the systems.  However, this complete discussion  requires an involved  mathematical
discussion which is out of the context of this  work.  However,  we will consider a simpler preliminary
analysis. It will be assumed  that the inclusion of the scalar field in
addition with matter, allows to  justify  that stability implies
the total mass of the  solution should grow when the initial condition for
the  density  of matter at the origin is also increased \cite{teukolsky}.

 This criterion can be easily checked  for the here obtained  solution,  by
considering the  total mass formula
\begin{align}
M(\epsilon_{0})  & =\int dr\text{ }4\pi\text{ }r^{2}\text{ }\epsilon_{t}(r),\\
\epsilon_{t}(r)  & =\frac{1}{2}(u(r)\phi^{\prime}{(r)}^{2}+\phi(r)^{2}%
+2j(r)\phi(r))+\epsilon(r),\nonumber
\end{align}
as a function of the  initial  condition for the matter energy density
$\epsilon(0)=\epsilon_{0}=40$ $p_{0}$. Note that the initial value of the
 scalar field at the origin $\phi_{0}$ is in fact  a function of
$\epsilon_{0}$,  which  value was found by imposing  the Yukawa behavior
of the  scalar field at large radius.  The new value of the total mass  after increasing  the matter
energy density at origin $\epsilon_{0}$, was then  derived   by solving the
equations for two nearly values of $\epsilon_{0}$,  the original one
$\epsilon_{0}=$2.3829  and the closer one  $\epsilon_{0}^{1}=$2.3929.
After modifying $\epsilon_{0}$ the new value of $ \phi_{0}$ was determined by
correspondingly modifying the old value in order to assure  the Yukawa
 dependence faraway from the origin.  Therefore,  the  approximate
evaluation of the derivative of the mass respect to the central density of
matter resulted in
\begin{align}
\frac{dM(\epsilon_{0})}{d\epsilon_{0}}  & \simeq\frac{M(\epsilon
_{0}+0.01)-M(\epsilon_{0})}{0.01}\nonumber\\
& =0.03504>0.
\end{align}

Thus,  the  obtained solution  satisfies this assumed stability criterion, which
is valid for standard stars being constituted only by fluids of matter.  As
mentioned  before,  the inclusion of the scalar field should imply more general
linear equations for the  oscillation modes of the system. However, it can be
expected that the considered here  criterion remains being valid.

 It should noted that the solution  derived is valid for any value of the
 scalar field mass. Thus, its stability criterion works for the  whole family of
 solutions for arbitrary  values of this mass.  It should be however,
still  investigated  the stability for the star configurations by varying
the initial condition for the energy density at the origin and the  intensity of the coupling of the matter with
 the scalar field. This will allow to determine whether
 there are  bounds for the  total mass  of the  obtained solutions.   We expect to
consider this  question in coming  works.

In ending this section, it should be remarked that the found static solution
had been allowed to exist thanks to the assumed interaction between the scalar
field and matter, as defined by the source of the field being proportional to
the matter density. It might be also helpful to note, that in reference
\cite{cb} the same type of interaction was employed to identify a static
Universe in which dark-energy interacts with matter. As well as it happened
here, the matter-scalar field interaction became central in the existence of
the static solution.

\section{The solution for photon-like matter.}

\ In this section we will consider the important special case in which the
assumed matter constitutive relation is associated to the photon-like gas.
That is, massless vector particles are assumed. They could be the real photons
or other dark-matter analogous massless vector particles. For these kind of
particles the traceless of the energy momentum tensor implies
\begin{equation}
\epsilon(r)=3\, p(r).
\end{equation}
\begin{figure}[h]
\begin{center}
\includegraphics[width=70mm]{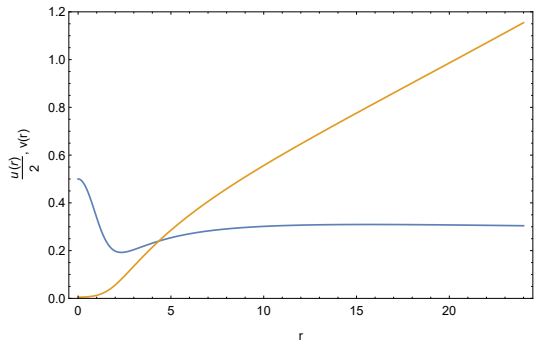} \label{phi1}
\end{center}
\caption{ The plots of the corresponding metric functions $u$ and $v$ with the
radial coordinates. They show how the gravitational potential, reflected by
$v$ grows nearly linearly with the coordinates at large radii. }%
\label{uv1}%
\end{figure}\begin{figure}[b]
\begin{center}
\includegraphics[width=70mm]{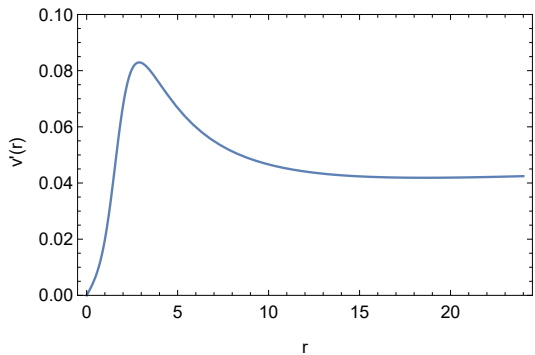} \label{phif}
\end{center}
\caption{ The radial dependence of the derivative of the function $v(r)$. This
quantity defines the gravitational force in the Newtonian limit. The behavior
indicates that the force over small massive bodies tends to be nearly constant
at large distances. }%
\label{force}%
\end{figure}

Since the considered particles are massless, it is possible that the solution
could show special characteristics, as it will be effectively the case.

The method of solving the EKG equations became identical to the one employed
for the case of nearly pressureless matter in the past section. The initial
conditions for the fields again were defined at the small radial distance
$\delta=10^{-6}$ in the form
\begin{align}
u(\delta)  &  =1,\\
v(\delta)  &  =1,\\
\phi(\delta)  &  =\phi_{0}=0.623048,\\
p(\delta)  &  =p_{0}=0.7943.
\end{align}

The interaction between the scalar field and the matter was fixed at the same
value employed in the past section%
\[
g=0.9.
\]

The EKG equations (\ref{eecuaadim1})-(\ref{eecuaadim4}) were then also solved
using the operation NDSolve of the programme Mathematica. The general
procedure for obtaining the solutions was exactly the same that was followed
in the past section. After that, the iterative process led to a definite
solution in which the scalar field also decreased at large radial values.
However, the net results became radically different from the physical point of
view, as it will be described in what follows.

\begin{figure}[h]
\begin{center}
\includegraphics[width=70mm]{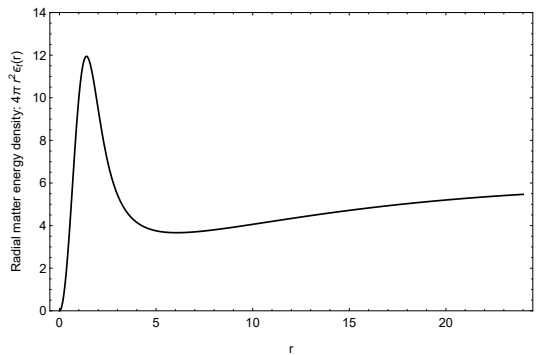}
\end{center}
\caption{The figure show the mass density by unit radial distance, as a
function of the radius, for the photon-like matter interacting with the
\ scalar field. It shows that the photons are able to escape to relatively
faraway regions from the symmetry center. }%
\label{radialmass}%
\end{figure}

The figure \ref{uv1} depicts the two metric components. As it can be observed,
at large distances from the symmetry center the field $v(r)$ in place of
becoming the constant Minkowski value, tends to get linear radial dependence.
This linear dependence implies that the gravitational field intensity of the
system is not decreasing with the distance. That is, the system is able to
produce an attractive nearly constant force on bodies situated faraway from
the symmetry center. The plot of this force $vs$ radius is shown in figure
\ref{force}. The appearance of this force means that the gravitation is unable
to trap the photon-like particles close to the centre. This effect is not out
of place to occur because the considered particles are massless, which move at
the velocity of light. This interpretation is clearly supported by the graph
in figure \ref{radialmass} which plots the total mass density up to a radial
distance $r$
\begin{align}
\rho_{t}(r)  &  =4\pi\text{ }r^{2}\epsilon_{t}(r),\\
\epsilon_{t}(r)  &  =\frac{1}{2}(u(r)\phi^{\prime}{(r)}^{2}+\phi
(r)^{2}+2j(r)\phi(r))+\epsilon(r).
\end{align}

The picture indicates that the photon-like energy at large distances is not
decaying, but slowly increases with the radius. It can remarked that this
photon-like energy is what effectively defines this effect, because the radial
decaying of the scalar field energy density makes the radial density of scalar
field energy ($4 \pi r^{2}$ times this energy density) to rapidly vanish with
the radius.

\begin{figure}[h]
\begin{center}
\hspace*{-0.7cm}\includegraphics[width=70mm]{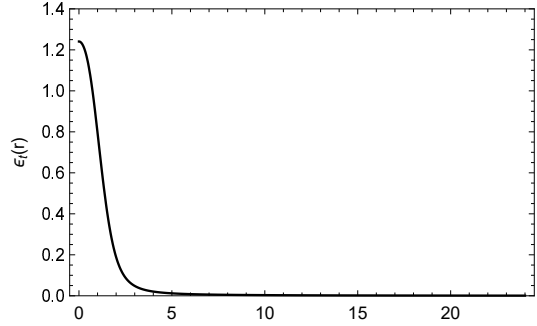} \vspace{-1cm}
\label{phif}
\end{center}
\caption{ The plot shows the total energy density as a function of the radial
distance for the photon-like matter interacting with a scalar field. Although
the density decreases with the radius, the mass accumulation does not stops as
the radius increases as it clear from figure \ref{radialmass}. }%
\label{phien}%
\end{figure}

The radial behavior of the scalar field is shown in figure \ref{phi1}. Far
from the center, the field again tends to rapidly decrease. \begin{figure}[h]
\begin{center}
\hspace*{-0.7cm}\includegraphics[width=70mm]{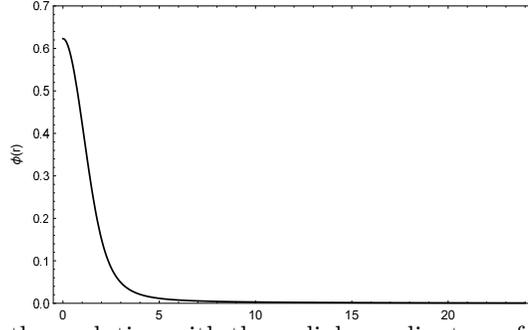} \vspace{-1cm}
\label{phif}
\end{center}
\caption{ The figure shows the evolution with the radial coordinate $r$ of the
scalar field for the photon-like matter-boson star.}%
\label{phi1}%
\end{figure}As for the matter density, for this alternative solution there is
a region close to the symmetry center showing the highest values of the
density. This vicinity is close to the one in which the scalar gets its
highest values. The radial dependence of the energy density is shown in figure
\ref{phien}. The vanishing of the energy density at large radii can give the
impression that the total mass of the structure is concentrated near the
origin. However, this shown to be not true, by the plot in figure
\ref{radialmass}.

In the following section, it will be argued that the found solution suggests
the possibility that photon-like matter can determine the velocity $vs$ radius
curves indicating the presence of dark-matter in the Universe.

\section{Dark-matter and photon-like-particles}

In this Section, it will be preliminarily examined the possibility of applying
the photon-like solutions to explain the dark-matter properties. For this
purpose, lets us assume that the considered solution is furnishing the
attraction required to define a rotational galaxy motion. In this situation
the circular equation of motion (in the original CGS units) of a given star of
mass $\delta m$, after assuming that its motion is defined by the
non-relativistic Newton equation, will be
\begin{equation}
-\delta m\frac{d}{d\varrho}\Phi_{G}(\varrho)=-\delta m\frac{V(\varrho)^{2}%
}{\varrho}. \label{newt}%
\end{equation}
\begin{figure}[h]
\begin{center}
\includegraphics[width=70mm]{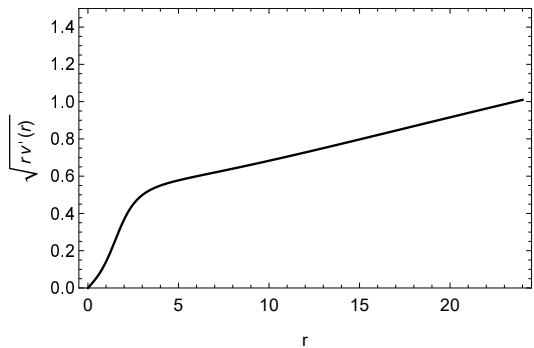}
\end{center}
\caption{The expression of the velocity $V(\rho)$ of the rotating star being
at radial distance $\rho$ as a function of the variable $r$.}%
\label{knee}%
\end{figure}That is, the gravitational force generated by the potential
$\Phi_{G}$ produces the centripetal acceleration of the circular motion. But,
assuming the Newtonian approximation for the gravitation potential it follows%

\begin{equation}
\Phi_{G}(\varrho)=\frac{c^{2}}{2}(v(\varrho)-1),
\end{equation}
which after substituted in (\ref{newt}) leads for the radial dependence of the
velocity the expression%
\begin{align}
V(\varrho)  &  =\sqrt{\frac{m\text{ }c^{2}}{2}\varrho\frac{dv(r)}{dr}}%
=\sqrt{\frac{c^{2}\text{ }r\text{ }v^{\prime}(r)}{2}}.\\
r  &  =m\text{ }\varrho,
\end{align}
where the derivative of $v$ over the radius $\varrho$ measured in cm has been
expressed in terms of derivative relative to the radial variables $r=m$
$\varrho$. Thus, the velocity at some radial distance $\varrho$ is only a
function of the $r$\ variable. This dependence is illustrated for the solution
being considered in the figure \ref{knee}.

It should be now recalled that we have found the solution for a particular
values of the energy density and scalar field at the origin of coordinates.
Therefore, the energy of the obtained field configuration is fixed, once its
parameters $m$ and $\kappa$ had been specified. Therefore the rotation curve
for the particular solution being investigated is only a function of $m$
(assumed the gravitational constant $\kappa$ is given). For illustrative
purposes, let us select a value of $m$ (the mass of the scalar field) such
that the radial position corresponding the knee in the curve shown in the
figure \ref{knee} at $r=$ $r_{c}=2.9469$ corresponds to a radial distance of
\begin{equation}
\varrho_{c}=7.517\text{ 10}^{3}\text{ ly}. \label{size}%
\end{equation}
This is a distance of the order of the sizes of the galaxies. Then, the value
of the scalar field mass turns to be
\begin{equation}
m=\frac{r_{c}}{\varrho_{c}}=4.1484\times10^{-22}\text{ \ gr. }%
\end{equation}

It should be noted that this is a mass for the boson of nearly one hundred
time larger than the proton mass. However, we have only chosen this value for
exemplifying the theoretical possibility that solutions of the considered form
could be associated to dark-matter. It is clear that many other possibilities
for the matter being coupled with the massless field can be considered in
substitution of the assumed scalar field. This type of field was relevant here
in connection with the first part of the work illustrating the existence of
static boson stars under coupling of this field with matter.

\begin{figure}[t]
\begin{center}
\includegraphics[width=70mm]{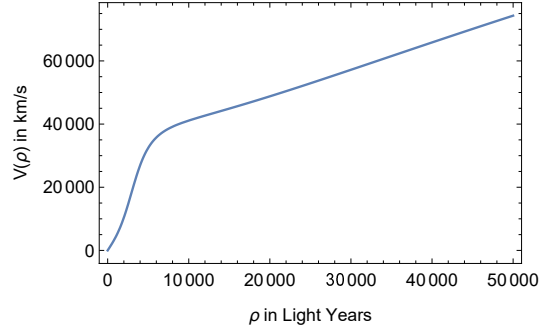}
\end{center}
\caption{ The figure shows the rotation curve for an hypothetical galaxy which
is generated by the photon-like matter in interaction with a scalar field for
the parameters chosen and the initial conditions for the matter at the origin
specified. The curve shows a type C of the three sorts (A, B and C) in the
classification of the galaxies according to the form of such curves
\cite{classif}. The type B corresponds almost constant velocities and the A
ones are associated to curves in that the velocities firstly rapidly increases
and then start diminishing almost linearly. The galaxy was defined by assuming
that the value $r_{c} $ of the variable $r$ at the "knee" of the curve in
figure \ref{knee} is defining a radial distance through $\rho_{c}=\frac{r_{c}%
}{m} $. }%
\label{rotcurve}%
\end{figure}

Further, the velocity $vs$ radius curve for the considered solution, assumed
that he scalar field mass is given by $m=4.1484\times10^{-22}$ \ gr , is shown
in figure \ref{rotcurve}. Note that the star velocities in km/s are smaller
than the light velocity, but not very much: only nearly an order of magnitude.
The considered solution defines a mass for the system which is larger than the
total mass of galaxies of the same size (by example Messier 33).

Therefore, in conclusion, it can be argued that particular forms of the
considered solution turn to be able in furnishing the total mass required for
galaxies of the observed types in Nature. It may be also expected that similar
effects can be obtained by coupling photon-like matter with the more usual
fermion matter.

\subsection{ Photon-like dark-matter}

It should be remarked that the currently most accepted view about the nature
of dark matter is that it is constituted by massive cold particles moving non
relativistically \cite{peebles,rees}. However, the dark matter problem is so
relevant that we estimate that it allows to explore all sort of ideas for
clarifying it. Since the here discussed solutions  indicates the possibility
of generating the complementary attractive forces acting in galaxies, below we
will consider some attention to examine the possibility that photon-like or
real photons could describe the dark matter properties.

\subsection{ A restriction for real photon matter}

Let us firstly consider a required condition for the solution examined here
(after assuming that it corresponds to real photons) can describe the galaxy
observations associated to dark-matter This requirement for photons to
generate dark-matter is that energy distribution of the photons should
approach the value associated to the CMB, at large distances from the centre
of the galaxy. This is a concrete requirement that can be easily checked for
the here obtained solution. By example, the CMB energy density in CGS units
has the formula and value
\begin{align}
\epsilon_{CMB}^{cgs}  &  =\frac{4}{c}\sigma T^{4}\nonumber\\
&  =6.1236\times10^{-13}\text{ \ }\frac{\text{erg}}{\text{cm}^{3}},\\
T  &  =3\text{ \ }^{0}K,
\end{align}
where $\sigma$ is the Stefan-Boltzman constant. But, expressed in the special
units defined in this work, it gets the value
\begin{align}
\epsilon_{CMB}  &  =\frac{8\pi l_{P}^{2}}{m^{2}}\epsilon_{CMB}^{cgs}%
\nonumber\\
&  =2.9417\times10^{-34}.
\end{align}
This result should be compared with the energy density of the considered
solution determining the galaxy size (\ref{size}) . This size after expressed
in the $r\ $spatial units used here gets the value
\begin{align*}
r  &  =m\text{ }\rho_{c}\\
&  =19.6016,
\end{align*}
where $m$, is the mass of the scalar field determining a solution with the
required total mass for generating a galaxy of similar size as the Messier 33
one. Therefore the energy density at distances from the symmetry center of the
order of the radius of the galaxy, takes the value
\begin{equation}
\epsilon(19.6016)=1.071\times10^{-3}>>\epsilon_{CMB}=2.9417\times10^{-34}.
\end{equation}

This result indicates that the photon like particles energy density of the
here investigated solution is enormously much larger than the CMB energy
density. Therefore, it allows to conclude that real photons are not able  to
describe dark-matter in the galaxy, if the solution considered is taken as it is.

However, there are still possibilities for overcoming the above obstacle. It
is clear that the growing potential continuously rising up to infinity, which
is generated in the considered solution is not observed in Nature. Therefore,
some effect should restrict this unlimited increasing. One possibility can be
that the photon dynamics could show a subtlety making the photon density to
decrease very much rapidly at the outside of the galaxy. This effect might be
related with the fact that the photon dynamics at any internal point of the
galaxy is being assumed as  the one valid for free photons: $\epsilon=3$ $p$.
But, the photons in the galaxy are subject to the "confining" gravitational
potential growing with the radial coordinates. Thus, the free photons can be
expected to be perturbed by the gravitational attraction. Let us in the
following subsection consider a simplified model to argue that such a
confining effect can in fact tend to trap the photons in some circumstances.

\subsection{Mechanisms for gravitationally bounding photons}

Let us consider the Maxwell equations in the presence of a special static
metric only depending on one Cartesian coordinates \cite{massive}%
\begin{align}
(\overrightarrow{\nabla}^{2}-\frac{\partial^{2}}{\partial x^{02}%
})\overrightarrow{E}(x)  &  =\overrightarrow{\nabla}(\overrightarrow{\nabla
}h(x).\overrightarrow{E}(x))-\frac{\partial}{\partial x^{0}}(\frac{\partial
}{\partial x^{0}}h(x)\overrightarrow{E}(x))-\nonumber\\
&  \overrightarrow{\nabla}(h(x))\times(\overrightarrow{\nabla}\times
\overrightarrow{E}(x)),\\
\overrightarrow{\nabla}\times\overrightarrow{E}(x)  &  =-\frac{\partial
}{\partial x^{0}}\overrightarrow{B}(x)
\end{align}
where $h(x)$ is a function of the determinant of the metric $g_{\mu\nu}$,
which in this subsection will be assumed to be
\begin{equation}
g_{\mu\nu}(x)=\left(
\begin{tabular}
[c]{llll}%
$g_{00}(x^{3})$ & $0$ & $0$ & $0$\\
$0$ & $1$ & $0$ & $0$\\
$0$ & $0$ & $1$ & $0$\\
$0$ & $0$ & $0$ & $1$%
\end{tabular}
, \ \right)
\end{equation}
in the space of the coordinates $\ x=(x^{0},x^{1},x^{2},x^{3}).$ That is, the
metric component $g_{00}(x^{3})$, which plays the role of the gravitational
potential in the Newtonian approximation, will be assumed to vary with the
coordinate $x^{3}.$ The function $h$ is given by given as
\begin{align}
h(x)  &  =\log(\sqrt[g]{-g(x^{3})})\nonumber\\
&  =\log(\sqrt{-g_{00}(x^{3})}),\\
\frac{\partial}{\partial x^{3}}h(x)  &  =\frac{1}{2}\frac{1}{g_{00}(x^{3}%
)}\frac{d}{dx^{3}}g_{00}(x^{3}).
\end{align}

Let us assume now that the metric has the form
\begin{align}
g_{00}(x^{3})  &  =-(1+\frac{2\Phi(x^{3})}{c^{2}})\nonumber\\
&  =-(1-2h\text{ }x^{3})
\end{align}
and the gravitational potential $\Phi(x^{3})$ is assumed to satisfies the
Newtonian approximation $\frac{\Phi(x^{3})}{c^{2}}<<1$ in a large region of
the coordinate $x^{3}.$ It corresponds with a Newtonian potential attracting
the masses to the negative $x^{3}$ axis. It is possible now to write the
simplifying equation for the considered constant potential%
\begin{equation}
\overrightarrow{\nabla}(h(x))=(0,0,-h).
\end{equation}

After the above definitions the Maxwell equations reduces to%
\begin{equation}
(\overrightarrow{\nabla}^{2}-\frac{\partial^{2}}{\partial x^{02}%
})\overrightarrow{E}(x)=-\overrightarrow{\nabla}(h(x))\times
(\overrightarrow{\nabla}\times\overrightarrow{E}(x)),
\end{equation}
where we have also used that we will search for waves defined by fields
$\overrightarrow{E}$ and $\overrightarrow{B}$ being orthogonal to the $x^{3}$
axis, and among themselves, which imply
\begin{equation}
\overrightarrow{\nabla}h(x).\overrightarrow{E}(x)=0,
\end{equation}

This constant form of the gradient of $h(x)$ allows the equation for the
electric field be simplified as follows%
\begin{equation}
(\overrightarrow{\nabla}^{2}-\frac{\partial^{2}}{\partial x^{02}%
})\overrightarrow{E}(x)=-h\frac{\partial}{\partial x^{3}}\overrightarrow{E}%
(x)).
\end{equation}
This systems of equations can be solved by \ Fourier transforming in the
variables \ $x^{0}$ and $\ x^{3}$. Then, expressing for the fields
\begin{equation}
\overrightarrow{E}(x)=(E,0,0)\exp(-\epsilon\text{ }x^{0}i+k^{3}x^{3}i),
\end{equation}
reduces the Maxwell equations to
\begin{equation}
(-(k^{3})^{2}+\epsilon^{2}+i\text{ }h\text{ }k^{3})\overrightarrow{E}(x)=0.
\end{equation}

Then, the waves should satisfy the dispersion relation%
\begin{equation}
k^{3}=\frac{h\text{ }i}{2}+\sqrt{\epsilon^{2}-\frac{h^{2}}{4}}.
\end{equation}

The explicit form of the solution for the electric field takes the form
\begin{align}
\overrightarrow{E}(x)  &  =(E,0,0)\exp(-\epsilon\text{ }x^{0}i+\sqrt
{\epsilon^{2}-\frac{h^{2}}{4}}x^{3}i)\times\nonumber\\
&  \exp(-\frac{h}{2}x^{3}).
\end{align}

This expression shows that the presence of the gravitational field introduces
a damping of the wave in the direction of its propagation, defined by the
positive $x^{3}$ axis. The damping is proportional to the  intensity of the
gravitational potential $h$. That is, the gravitational field affects the
dynamics of the photons tending to trap them in the region of lower potential values.

This effect, strongly suggests that the gravitational field can play a role
determining that found solutions for photon-like matter, after corrected for
the photon dynamics, can show metrics tending to the Minkowski one for large
values of the radius. It this happens, it becomes an open question again
whether such corrected solutions could or not show photon energy densities
being able to reduce at large radii to the value associated to the CMB. This
was the main obstacle posed before for the photons behave as dark-matter.

In this sense, we would like to formulate a task that could help in discussing
the above issue. It can be defined as consisting in formulating a classical
statistical descriptions for massless particles, but subject to the local
metric in each interior point of a gravitational field region. Such an
analysis can correct for the neglected influence of the gravitational field
over the employed free photon dispersion relation leading to the $\epsilon=3$
$p$.

In the next subsection we will also remark about an effect that can support
that photons can produce the effects of the dark-matter.

\subsection{ The Tolman theorem}

If the obstacles noted above can be surmounted, one important concept can
play a role in the discussion of the currently mostly rejected possibility
that photon can constitute the dark-matter. It is the so called Tolman theorem
\cite{tolman}, that states that the temperature of radiation filling in
thermal equilibrium a spatial region in which a static gravitational potential
is acting, is scaled with the value of the temporal component of the metric
$g_{00}$ as
\begin{equation}
T(x)=\frac{T_{\infty}}{g_{00}(\overrightarrow{x})},
\end{equation}
where $x=(c\,t,\overrightarrow{x})$ are the space-time coordinates of a point
and $T_{\infty}$is the temperature in the assumed faraway spatial Minkowski
like regions. Note that in this subsection we have returned to the positive
signature metric for $g_{00}.$

This effect has to do with the possibility that real photon matter can be
observed from the Earth, if playing the examined role of dark-matter. It is
true, that the considered solution retains its interest if the massless
particles generating it are sorts of really dark-matter particle having a
photon-like massless vector nature. However, such solutions could be even more
motivating if the usual photon can be responsible for creating the dark-matter
effects after surpassing the already posed limitations.

For the situation under consideration, the natural setting for the temperature
in the regions far from the galaxy is the CMB radiation temperature of nearly
3 K$^{0}$. Therefore, the radiation temperature at the interior point of the
galaxy can take very much increased temperatures than 3 K$^{0}$ as the
gravitational potential reduces near the center. However, the arriving to the
Earth radiation coming from an arbitrary interior point of the galaxy, will
always have a frequency spectrum in the microwave region. This is because the
light should "climb" the gravitational potential barrier, which reduces its
frequency spectrum down again to the CMB one. Thus, since the CMB associated
frequencies are very far form the observable light spectrum, in this first
instance, the real photons might play the role of dark matter. However, it
should be recalled that in order to perform this function, it is required
to explain how the photon gas at the interior of the structure can become in
equilibrium with the CMB at the galaxies exterior regions. At first sight,
this possibility seems to be difficult, due to the drastic lack of balance
with the CMB radiation which was evaluated before for the solution presented here.

\section*{Summary}

The role of the interaction between a real scalar field and matter in the
solutions of the EKG equations is investigated \cite{jetzer,liddle1,liddle2,
urena}. Stars showing a static scalar field are argued to exist thanks to the
interaction between the scalar field and matter. The field, matter and density
distributions are evaluated as smooth functions of the radial distance to the
symmetry center. The solutions satisfy a standard stability criterion which is obeyed  by
simpler stars constituted only by fluids \cite{teukolsky}. However,  to conclude their stable character, a  closer investigation should be done of the more complex linearized equations for the oscillation modes, which include  a scalar field in addition to a fluid.

The case of photon like matter is also examined.
Surprisingly, in this case there exist star like solutions for which the
radial gravitational potential at large distances grows linearly. This effect
is related with the massless character of the particles being considered. The
emerging behavior leads to suspect that it could be helpful for the
understanding the origin of the dark-matter effects. In a preliminary
examination of this question, we determine the predictions of the solutions
for the rotation curves of a galaxy. The parameters of the solution are chosen
to define a galaxy of size being similar to the ones observed. The form of the
velocity curve obtained corresponds to a C type of galaxies according to a
standard classification \cite{classif}. It is started to be explored the
possibility that  correcting the free photon constitutive relation $\epsilon=3
p$ for the effects of the gravitational  field may stop the linear growing of
the gravitational potential obtained here. A simple model  is solved to
inspect the action of the Newtonian gravitational potential over a plane wave
propagating against it.  The solutions supports that the inclusion of the
metric in the classical photon statistics can produce a bounding effect over
the photons. A model is then just formulated for to consider the corrected
statistics. The idea is to derive the constitutive relations for photons
moving at the light velocity, but in the space-time dependent metric. This
problem is expected to be considered elsewhere.

It is also discussed whether or not a gas of real photons can play the role of
dark-matter, or it should be described by photon like real dark-matter
particles. For this purpose the evaluation of the energy density of the photon
like constitutive matter at the regions outside of the modeled galaxy was
evaluated. The result was very much larger than the CMB corresponding value.
This evaluation does not support the identification with photons of the photon
like matter determining the rotation curves, if the solution is taken as it
is. However, the determination of a corrected solution  after improving the
photon like dynamics, as proposed before, can still allow to satisfy this criterion.

It was also presented an observation based in the Tolman theorem. This theorem
defines the temperature of the photon radiation assumed to determine the
solution considered here, as the CMB temperature after divided the temporal
component of the metric. One important property of the theorem, gives a
partial support to photons as creating the dark matter potential.  It is the
fact that, no matter the high the temperature of the photons will be at an
interior point of the galaxy, the light coming to the Earth from this point
will always arrive with a $3\, ^{0}K $ frequency spectrum, coinciding with the
CMB radiation one, coming from any other direction.

There remains an important issue to be further addressed in connection with
the found solutions. It is the question about their stability. For this
purpose in a coming study, the spectrum of the linearized equations of motion
for small radial perturbations will be investigated.

\section*{Acknowledgments}

The authors very much thank the support received from the Office of External
Activities of ICTP (OEA), through the Network on \textit{Quantum Mechanics,
Particles and Fields} (Net-09).

\end{document}